\documentclass[twocolumn,preprintnumbers,fleqn,amsmath,amssymb,showkeysX]{revtex4-1}

\usepackage{multirow}
\usepackage{lipsum}
\usepackage{siunitx}
\usepackage{graphicx,epsfig,epstopdf}
\usepackage{bm}
\usepackage{epstopdf}
\usepackage[a4paper,margin=1 cm]{geometry}
\usepackage{amsmath}
\usepackage{float}
\usepackage{breqn}
\usepackage{slashed}
\usepackage{amssymb}
\usepackage{gensymb}
\usepackage{simplewick}

\makeatletter
\let\cat@comma@active\@empty   
\makeatother
\makeatletter
\newcommand*{\rom}[1]{\expandafter\@slowromancap\romannumeral #1@}
\makeatother

\begin{document}
\title{Eta-nucleon and eta-prime-nucleon coupling constants in QCD and the role of gluons}

\author{Janardan P. Singh}
\email{janardanmsu@yahoo.com, retired from services of The Maharaja Sayajirao University of Baroda, Vadodara}
\affiliation{Physics Department, Faculty of Science, The Maharaja Sayajirao University of Baroda, Vadodara 390002, Gujarat, INDIA}

\author{Shesha D. Patel}
\email{sheshapatel30@gmail.com}
\affiliation{Physics Department, Faculty of Science, The Maharaja Sayajirao University of Baroda, Vadodara 390002, Gujarat, INDIA}

\date{\today}

\begin{abstract}
Coupling constants of $\eta$ and $\eta^{\prime}$ mesons with nucleons have been calculated using the method of QCD sum rules. Starting from vacuum-to-meson correlation function of interpolating fields of two nucleons, its matrix element with respect to nucleon spinors has been considered. Coupling constants at the physical points have been estimated from extrapolation of results obtained at two other points. Anomalous glue has been found to give substantial contribution to the coupling constants and also accounts for a significant OZI-rule violation.
\end{abstract}

\maketitle
\section{Introduction}
Study of $\eta$-$\eta^{\prime}$ mesons reveals interesting interplay between chiral symmetry breaking and gluon dynamics. The fact that $\eta$ and $\eta^{\prime}$ mesons have masses much higher than the values they would have if these mesons were merely Nambu-Goldstone bosons is a result of their association with non-perturbative gluon dynamics and QCD axial anomaly\cite{Witten:1979}. While some authors from the study of $\eta^{(\prime)}$  transition form factor have concluded quite large (radiatively generated) two gluon Fock components in $\eta^{\prime}$ wavefunction\cite{Kroll:2013}, others, including one of us, have contended that at low energies $(Q^{2}\approx2\mathrm{{GeV}^{2}})$ only about 20 \% of the momentum of $\eta$ and $\eta^{\prime}$ mesons is carried by gluons in the chiral limit, though at very high energies $\thicksim M_{z}^{2}$ this fraction evolves to 50\% \cite{Singh:2012}. In case of nucleons, roughly 50\% of the momentum is carried by gluons at moderately high energies. This conclusion on momentum fraction carried by gluons is drawn from results on valance quark distribution functions of the hadron following conventional logic. A claim of a significant glue content of $\eta^{\prime}$ by KLOE collaboration \cite{Amb:2009} is based on the fitting measurements on $\phi\rightarrow \eta^{\prime}\gamma,\eta\gamma$ together with several radiative decays, such as $V\rightarrow P\gamma$ and  $P\rightarrow V\gamma$ involving $\eta^{\prime}$, $\eta$ mesons. This is in conflict with the analysis of radiative decays $V\rightarrow P\gamma$ and  $P\rightarrow V\gamma$ by Escribano and Nadal \cite{Esc:2007} and Thomas \cite{Thomas:2007} who found no evidence for gluonic contribution in $\eta$ or $\eta^{\prime}$.
\par $\eta^{(\prime)}$ -phenomenology is characterized by large OZI-violations\cite{Bass:1999}. The flavor-singlet Goldberger-Trieman relation relates $\eta^{\prime}$-nucleon coupling constant to the flavor-singlet axial charge of the nucleon $g_{A}^{(0)}$ extracted from polarized deep inelastic scattering\cite{Shore:1992,Shore:2006}. Harland-Lang et al. \cite{Har:2013} have considered the central exclusive production of $\eta^{\prime}$, $\eta$ meson pairs in pp($\overline{p}$) collision in the perturbative regime. They show on phenomenological ground that the cross-sections for the production of $\eta$, $\eta^{\prime}$ meson pairs in such processes are strongly sensitive to the size of the gluon content of these mesons. Bass and Moskal \cite{Bass:2018} have emphasized the fact that the magnitude of the scattering lengths $a_{\eta N}$ and $a_{\eta^{\prime}N}$ are much greater than $a_{\pi N}$. They have further pointed out that the search for $\eta$ and $\eta^{\prime}$ mesic nuclei will help pin down the dynamics of axial U(1) symmetry breaking in low-energy QCD. Gluons in the proton play an essential role in understanding its internal spin structure\cite{Bass:2007}. A reliable determination of the coupling constants $g_{\eta NN}$ and $g_{\eta^{\prime} NN}$ would shed considerable light on the axial $U(1)$ dynamics of QCD\cite{Shore:2006}. Also, $g_{\eta NN}$ and $g_{\eta^{\prime} NN}$ have essential roles in construction of realistic NN potential\cite{Rijken:2006,Machleidt:2001}, in estimates of electric dipole moment of the neutron\cite{Kawarabayashi:1980} and in analyses of photoproduction of these mesons\cite{Tiator:2018,Nys:2017,Aznauryan:2003} and scatterings involving nucleons and these mesons. However, measurement of these coupling constants is a formidable task: the production of $\eta$ mesons from a single nucleon is dominated by resonance $S_{11}(1535)$\cite{Shore:2006} and there is a strong indication of four nucleon resonances with masses close to and above $\eta^{\prime}N$ threshold contributing to $\eta^{\prime}$ production\cite{Anisovich:2017}. Therefore, a reliable theoretical estimate of $g_{\eta NN}$ and $g_{\eta^{\prime} NN}$ is desirable.  QCD sum rules have been used in past to calculate $\eta$-nucleon coupling constant $g_{\eta NN}$ in SU(3) symmetry limit\cite{Kim:2000} as well as with SU(3)- flavor violating effects taken into account\cite{Singh:2011}; it has also been used to calculate singlet axial-vector coupling constant of the nucleon without resorting to any instanton contribution \cite{Singh:2015}. In Ref\cite{Singh:2011}, $\eta$ was considered as a member of octet family $\eta_{8}$ and no mixing with the singlet $\eta_{0}$ was taken into account for calculating $g_{\eta NN}$.
\par In the present work, we calculate the coupling constants of both physical $\eta$ and $\eta^{\prime}$ mesons with a nucleon using quark-flavor basis which turns out to be more appropriate for working with a singlet component. Light-cone expansion of a quark propagator naturally gives emission of anomalous gluons which couple to $\eta$ and $\eta^{\prime}$ mesons. This is a characteristic contribution to $g_{\eta NN}$ and $g_{\eta^{\prime} NN}$ having implication for nucleon spin problem\cite{Bass:2007}.

\section{Formalism and Construction of Sum Rule}
General methods of calculating meson-nucleon coupling constants have been developed in Refs. \cite{Reinders:1985,Birse:1996,Kim:2000,Kondo:2003,Doi:2004}. Consider the correlator of the nucleon current between vacuum and one $\eta^{(\prime)}$- state,
\begin{equation}\label{eq1}
\Pi(q,p)=i\int d^{4}x e^{iqx}\langle0\mid T\{J_{N}(x),\overline{J_{N}}(0)\}\mid\eta^{(\prime)}(p)\rangle,
\end{equation}
where $J_{N}$ is the standard proton current\cite{Ioffe:2010}.
\begin{dmath}\label{eq2}
J_{N}=\epsilon^{abc}[u^{aT}C\gamma_{\mu}u^{b}]\gamma_{5}\gamma^{\mu}d^{c},
\end{dmath}
Here a,b,c are color indices. The $\eta^{(\prime)} NN$ coupling constant $g_{\eta^{(\prime)} NN}$ is defined through the coefficient of the pole as\cite{Kondo:2003}:
\begin{dmath}\label{eq3}
{\footnotesize \overline{u}(qr)(\slashed{q}-M_{n})\Pi(q,p)(\slashed{q}-\slashed{p}-M_{n})u(ks)\mid_{q^{2}= M_{n}^{2},(q-p)^{2}=M_{n}^{2}}}= i\lambda^{2}g_{\eta^{(\prime)} NN}\overline{u}(qr)\gamma_{5}u(ks),
\end{dmath}
where $k=q-p$, u(qr) is a Dirac spinor and $\lambda$ is the coupling constant of the proton current with one-proton state\cite{Ioffe:2010}.
\begin{dmath}\label{eq4}
\langle0\mid J_{N}(0)\mid q\rangle =\lambda u(q).
\end{dmath}
Following Ref.\cite{Kondo:2003} we define the projected correlation function
\begin{dmath}\label{eq5}
\Pi_{+}(q,p)=\overline{u}(qr)\gamma_{0}\Pi(q,p)\gamma_{0}u(ks).
\end{dmath}
$\Pi_{+}$ can be regarded as a function of $q_{0}$ in the reference frame in which $\bf{q}=0$. The even and odd parts of the $\Pi_{+}$ satisfy dispersion relations as,
\begin{eqnarray}\label{eq6}
\Pi_{+}^{E}(q_{0}^{2})&=&-\frac{1}{\pi}\int dq_{0}^{\prime}\frac{q_{0}^{\prime}}{q_{0}^{2}-q_{0}^{\prime 2}} \text{Im} \Pi_{+}(q_{0}^{\prime}), \nonumber\\
\Pi_{+}^{O}(q_{0}^{2})&=&-\frac{1}{\pi}\int dq_{0}^{\prime}\frac{1}{q_{0}^{2}-q_{0}^{\prime 2}} \text{Im} \Pi_{+}(q_{0}^{\prime}).
\end{eqnarray}
On taking Borel transform\cite{Kondo:2003,Ioffe:2010} with respect to $q_{0}^{2}$ they take the form
\begin{eqnarray}\label{eq7}
\hat{B}[\Pi_{+}^{E}(q_{0}^{2})]&=&\frac{1}{\pi}\int d q_{0}^{\prime}q_{0}^{\prime} e^{\frac{-{q_{0}^{\prime}}^{2}}{M^{2}}} \text{Im} \Pi_{+}(q_{0}^{\prime}),  \nonumber \\
\hat{B}[\Pi_{+}^{O}(q_{0}^{2})]&=&\frac{1}{\pi}\int d q_{0}^{\prime} e^{\frac{-{q_{0}^{\prime}}^{2}}{M^{2}}} \text{Im} \Pi_{+}(q_{0}^{\prime}),
\end{eqnarray}
where M is the Borel mass parameter. The RHS of Eq. (\ref{eq7}) is  expanded in terms of the observed spectral function. The absorptive part of the projected correlation function can be written as,
\begin{dmath}\label{eq8}
\text{Im}\Pi_{+}(q,p)=-\overline{u}(qr)i\gamma_{5}u(ks)\pi\lambda^{2}g(q_{0},{\bf{p}}^{2}) \Big\{\frac{\delta(q_{0}-M_{n})}{q_{0}-E_{k}-\omega_{p}}+\frac{\delta(q_{0}-E_{k}-\omega_{p})}{q_{0}-M_{n}}\Big\}+ \Big\{\theta(q_{0}-s_{\eta^{(\prime)}})+\theta(-q_{0}-s_{\eta^{(\prime)}})\Big\}\text{Im}\Pi_{+}^{OPE}(q,p),
\end{dmath}
where $s_{\eta^{(\prime)}}$ is the effective continuum threshold.
\par Before we start QCD calculation of the correlator, we introduce notations of quantities which appear in quark-flavor basis. One introduces quark states \cite{Feldmann:2000}
\begin{equation}
\mid \eta_{q}\rangle=\frac{1}{\sqrt{2}}\mid(u\overline{u}+d\overline{d})\rangle,
\mid \eta_{s}\rangle=\mid s\overline{s}\rangle,
\end{equation}
in terms of which physical meson states are written as follows:
\begin{equation}
\left(\begin{array}{c}\eta\\\eta^{\prime}\\ \end{array}\right)=U(\phi)\left(\begin{array}{c}\eta_{q}\\\eta_{s}\\ \end{array}\right),
U(\phi)=\left(\begin{array}{cc}\cos\phi &-\sin\phi \\\sin\phi &\cos\phi\\ \end{array}\right).
\end{equation}
In this basis decay constants $f_{M}^{(q)}$ and $f_{M}^{(s)}$ for a meson $M=\eta,\eta^{\prime}$ are defined as
\begin{eqnarray}
\langle 0\mid \frac{1}{\sqrt{2}}\{\overline{u}(0)\gamma_{\mu}\gamma_{5}u(0)+\overline{d}(0)\gamma_{\mu}\gamma_{5}d(0)\}\mid M(p)\rangle &=& if_{M}^{(q)}p_{\mu},\nonumber\\
\langle 0\mid \overline{s}(0)\gamma _{\mu}\gamma_{5}s(0)\mid M(p)\rangle =i f_{M}^{(s)}p_{\mu}.
\end{eqnarray}
These decay constants are parameterized as
\begin{dmath}\label{eqss}
\left(\begin{array}{cc}f_{\eta}^{(q)} & f_{\eta}^{(s)}\\f_{{\eta}^{\prime}}^{(q)} & f_{{\eta}^{\prime}}^{(s)}\\ \end{array}\right)=U(\phi)\left(\begin{array}{cc}f_{q} & 0  \\0 & f_{s}\\ \end{array}\right).
\end{dmath}
Matrix elements of pseudoscalar currents between vacuum and one-meson states introduce another set of constants $h_{M}^{(q)}$, $h_{M}^{(s)}$:
\begin{eqnarray}
2im_{q}\langle 0\mid \frac{1}{\sqrt{2}}\{\overline{u}(0)\gamma_{5}u(0)+\overline{d}(0)\gamma_{5}d(0)\}\mid M(p)\rangle&=&h_{M}^{(q)},\nonumber\\
2im_{s}\langle 0\mid \overline{s}(0)\gamma_{5}s(0)\mid M(p)\rangle&=&h_{M}^{(s)}.
\end{eqnarray}
These constants are also parameterized through a relation similar to Eq. (\ref{eqss}).
\par In QCD, the correlator is calculated via the OPE at the deep space-like region $q^{2}\rightarrow-\infty$. In the present approach, using vacuum saturation hypothesis, the quark-antiquark component with the $\eta^{(\prime)}$ meson is factored out from the correlator. The rest of the correlator is time-ordered products of quark fields \cite{Kim:2000}. This is a light-cone expansion of the correlation function which is the first step in calculating Wilson coefficients of short-distance expansion (SDE). In the second step, short-distance expansion of light-cone operators is performed \cite{Kondo:2003,Doi:2004}. Additionally, contribution to the correlator arising from radiatively generated anomalous gluons from quark propagators has also been taken into account. There are three kinds of nonlocal bilinear quark operators which contribute to the vacuum-to-meson matrix elements as given below. While the results on SDE of the matrix element of axial-vector type nonlocal quark operators have been taken from Ref. \cite{Kim:2000}, the other two results, namely, those concerning pseudoscalar and tensor type nonlocal quark operators, have been calculated by us basing on parametrization given in Refs. \cite{Belyaev:1995,Ball:1999}. Finally, quark-gluon mixed contribution to the matrix element are obtained by moving a gluon field-strength tensor from a quark propagator into the quark-antiquark  component with a $\eta^{(\prime)}$ meson. Our calculation of this matrix element is based on results and parametrization of Refs. \cite{Doi:2004,Ball:1999}. We list below the results of the vacuum-to-$\eta^{(\prime)}$ matrix elements of the light-cone operators used in this paper $(q=u,d)$:
\begin{eqnarray}\label{eq9}
&\langle & 0\mid\overline{q}(0)i\gamma_{5}q(x)\mid\eta^{(\prime)}(p)\rangle \cr && = \frac{h_{q}}{2\sqrt{2}m_{q}}\binom{C}{S}\Big\{1-\frac{i p\cdot x}{2}-\frac{(p\cdot x)^{2}}{6}+\frac{i(p\cdot x)^{3}}{24}\Big\}, \nonumber \\ \nonumber \\
&\langle & 0\mid\overline{q}(0)\gamma_{\mu}\gamma_{5}q(x)\mid\eta^{(\prime)}(p)\rangle \cr &&=\frac{f_{q}}{\sqrt{2}}\binom{C}{S}\Big\{1-\frac{i p\cdot x}{2}\Big\}\Big\{i p_{\mu}-\frac{i\delta^{2}p\cdot x x_{\mu}}{18}+\frac{5i\delta^{2}x^{2}p_{\mu}}{36}\Big\}, \nonumber \\ \nonumber \\
&\langle & 0\mid\bar{q}(0)\gamma_{5}\sigma^{\mu\nu}q(x)\mid\eta^{(\prime)}(p)\rangle \cr && =i\frac{h_{q}}{12 \sqrt{2} m_{q}}\binom{C}{S}\Big\{p^{\mu}x^{\nu}-p^{\nu}x^{\mu}\Big\} \cr && \times \Big\{1-\frac{i p\cdot x}{2}-\frac{3 (p\cdot x)^{2}}{20}+\frac{i(p\cdot x)^{3}}{30}\Big\}, \nonumber \\ \nonumber \\
&\langle& 0\mid q^{a}(x)g G^{n}_{\mu\nu}(\frac{x}{2})\overline{q}^{b}(0)\mid\eta^{(\prime)}(p)\rangle \cr && = i\binom{C}{S}\frac{(t^{n})_{ab}}{16\sqrt{2}} \Big\{f_{q}\gamma_{5}a_{1}(p_{\nu}\gamma_{\mu}-p_{\mu}\gamma_{\nu})p\cdot x \cr && + a_2 \slashed{p} (p_{\mu}x_{\nu}-p_{\nu}x_{\mu})-\cr && f_{3\pi}\gamma_{5}p^{\alpha}(p_{\mu}\sigma_{\alpha\nu}-p_{\nu}\sigma_{\alpha\mu})(1-\frac{ip\cdot x}{2}) \cr && - f_{q}\epsilon_{\mu\nu\alpha\beta}\gamma^{\alpha}p^{\beta}\frac{\delta^{2}}{3}(1-\frac{ip\cdot x}{2})\Big\},
\end{eqnarray}
where
$\binom{C}{S} =\binom{\cos\phi}{\sin\phi}$,  $a_{1}=\binom{0.0074}{0.0062}$, $a_{2}=\binom{0.0136}{0.0149}$  for $\binom{\eta}{\eta^{\prime}}$, $\phi$ is mixing angle in quark-flavor scheme\cite{Feldmann:2000}.
The relevant gluonic term can be extracted from the light- cone expansion of the quark propagator \cite{Balitsky:1989}:
\begin{widetext}
\begin{dmath}\label{eq10}
\langle0\mid T\{q(x)\overline{q}(0)\}\mid 0 \rangle =\frac{\Gamma(-\epsilon)}{32\pi^{2}(-x^{2})^{-\epsilon}}x^{\mu}\gamma_{\lambda}\gamma_{5}g^{2}\int_{0}^{1}
du\int_{0}^{u}dv[(1-2\overline{u}-2v)G_{\mu\nu}(ux)\tilde{{G}^{\nu}}_{\lambda}(vx)-\tilde{G}_{\mu\nu}(ux){G^{\nu}}_{\lambda}(vx)]+.......\rightarrow
\frac{\Gamma(-\epsilon)}{32\pi^{2}(-x^{2})^{-\epsilon}}g^{2}x^{\mu}\gamma_{\lambda}\gamma_{5}t^{a}t^{b}{G^{a}_{\mu}}^{\nu}(0)G^{b}_{\nu\lambda}(0)\int_{0}^{1}du\int_{0}^{u}dv(-2\overline{u}-2v)+.....\rightarrow
\frac{\Gamma(-\epsilon)\alpha_{s}}{72\times4\pi(-x^{2})^{-\epsilon}}G\widetilde{G}\slashed{x}\gamma_{5}+......
\end{dmath}
\end{widetext}
for dimension $d$=$4-2\epsilon$, $G\widetilde{G}$ is evaluated at the origin in the spirit of short distance expansion and ellipsis stand for other structures. Here $G\widetilde{G}=\frac{1}{2}\epsilon^{\mu\nu\rho\sigma}G_{{\mu\nu}}^{a}G_{{\rho\sigma}}^{a}, \epsilon^{0123}=+1$.
Defining matrix element \cite{Beneke:2003},
\begin{dmath}\label{eq11}
\langle 0\mid \frac{\alpha_{s}}{4\pi}G\tilde{G}\mid\eta^{(\prime)}(p)\rangle=a_{\eta^{(\prime)}},
\end{dmath}
we can write:
\begin{dmath}\label{eq12}
\langle 0\mid T\{q(x)\overline{q}(0)\}\mid\eta^{(\prime)}\rangle =\frac{\Gamma(-\epsilon)}{72}(-x^{2})^{\epsilon}a_{\eta^{(\prime)}}\slashed{x}\gamma_{5}+..........
\end{dmath}
where
\begin{eqnarray}\label{eq13}
a_{\eta}&=&-\frac{m_{\eta^{\prime}}^{2}-m_{\eta}^{2}}{\sqrt{2}}\sin\phi\cos\phi(-f_{q}\sin\phi+\sqrt{2}f_{s}\cos\phi), \nonumber\\
a_{\eta^{\prime}}&=&-\frac{m_{\eta^{\prime}}^{2}-m_{\eta}^{2}}{\sqrt{2}}\sin\phi\cos\phi(f_{q}\cos\phi+\sqrt{2}f_{s}\sin\phi).
\end{eqnarray}
The removal of divergence requires renormalization and we do this in MS scheme. In $\Pi(q,p)$ the operator $G\tilde{G}$ originates from only d-quark line and there is no contribution where gluons originate from two different quark lines.
\begin{widetext}
\begin{figure}[H]\centering
\includegraphics[width=0.6\textwidth]{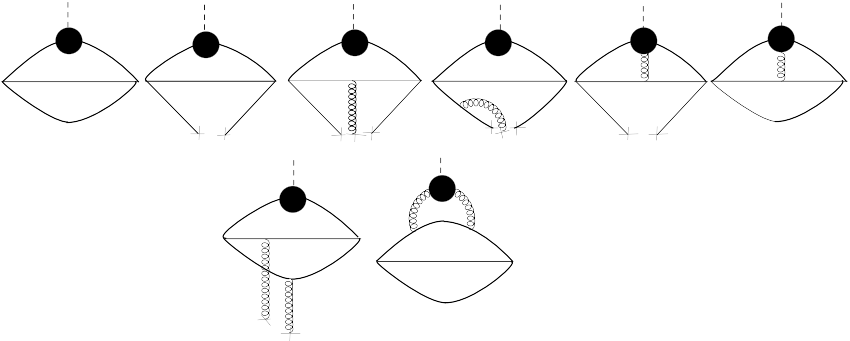}
\caption{Feynman diagrams considered for the correlation function $\Pi(q,p)$.The dotted line stands for the meson.}
\end{figure}
\end{widetext}
We get the following result for the correlator:
\begin{widetext}
\begin{eqnarray}\label{eq14}
\Pi(q,p)\cr && =
\frac{f_{q}}{\sqrt{2}}\binom{C}{S}\Big\{\frac{i}{6\pi^{2}}\Big[\slashed{p}
\Big(\log{(-q^{2})}\Big\{q^{2}-2q\cdot p-\delta^{2}\Big\}+\frac{\delta^{2}2 q\cdot p}{q^{2}}-3\log{(-q^{2})}
\Big\{q^{2}-q\cdot p-\delta^{2}\Big\}-\frac{8\delta^{2}q\cdot p}{3 q^{2}}\Big)+ \cr &&
\slashed{q}\Big(\log{(-q^{2})}\Big\{2q\cdot p-p^{2}\Big\}-\frac{2(q\cdot p)^{2}}{q^{2}}+\delta^{2}
\Big\{\frac{p^{2}}{q^{2}}- \frac{2(p\cdot q)^{2}}{q^{4}}\Big\}-\frac{\delta^{2}2q\cdot p}{q^{2}}-
\frac{2\delta^{2}}{3}\Big\{\frac{q\cdot p}{q^{2}}-\frac{p^{2}}{2 q^{2}}+
\frac{(q\cdot p)^{2}}{q^{4}}\Big\}\Big)\Big]\gamma_{5}+ \cr && \frac{i\langle g^{2}G^{2}\rangle}{144\pi^{2}}
\Big[\slashed{p}\Big(-\frac{4}{q^{2}}-\frac{3q\cdot p}{q^{4}}+\frac{8\delta^{2}}{9 q^{4}}\Big)+
\slashed{q}\Big(\frac{p^{2}-2q\cdot p}{q^{4}}-\frac{4(q\cdot p)^{2}}{q^{6}}-\frac{32\delta^{2}q\cdot p}{9q^{6}}\Big)\Big]\gamma_{5}- \frac{i\delta^{2}}{18\pi^{2}}\Big[\slashed{p}\log{(-q^{2})}+
\frac{2\slashed{q}q\cdot p}{q^{2}}- \cr && \frac{\slashed{q}p^{2}}{q^{2}}-
\frac{\slashed{p}q\cdot p}{q^{2}}+\frac{2\slashed{q}(q\cdot p)^{2}}{q^{4}}\Big]\gamma_{5}\Big\} -\frac{i\slashed{q}\gamma_{5}a_{\eta^{(\prime)}}\log{(\frac{-q^{2}}{\mu^{2}})}}{6\pi^{2}}\Big\{3+\log{(4)}-2\gamma-\frac{1}{2}
\log{\Big(\frac{-q^{2}}{\mu^{2}}\Big)}\Big\}+\frac{ih_{q}\langle\overline{q}q\rangle}{18\sqrt{2}m_{q}}\binom{C}{S} \times \cr && \Big\{\frac{2\slashed{p}}{q^{2}}+\frac{1}{q^{4}}\Big[4\slashed{q}p\cdot q-2\slashed{q}p^{2}+ \frac{8\slashed{q}\Big(q\cdot p\Big)^{2}}{q^{2}}-\frac{3}{5}\Big(-\slashed{p}p^{2}+\frac{12}{q^{2}}\Big\{\slashed{q}p^{2}p\cdot q+\frac{\slashed{p}(p\cdot q)^{2}}{3}-\frac{2\slashed{q}(p\cdot q)^{3}}{q^{4}}\Big\}\Big)\Big]\Big\}\gamma_{5}-\frac{ih_{q}\langle\overline{q}g\sigma\cdot Gq\rangle}{24\sqrt{2}m_{q}q^{4}}\times \cr &&\binom{C}{S}\Big\{\slashed{p}-\frac{4\slashed{q}q\cdot p}{q^{2}}\Big\}\gamma_{5}  +\frac{h_{q}i\gamma_{5}}{8\pi^{2}\sqrt{2}m_{q}} \times \binom{C}{S}\Big\{\log{(-q^{2})}(q^{2}-q\cdot p)+\frac{1}{3}\Big[p^{2}\log{(-q^{2})}+\frac{2(q\cdot p)^{2}}{q^{2}}\Big]-\cr &&\frac{1}{3}\Big[\frac{3 p^{2}p\cdot q}{2 q^{2}}-\frac{(p\cdot q)^{3}}{q^{4}}\Big]\Big\}-\langle g^{2}G^{2}\rangle \frac{i\gamma_{5}h_{q}}{192\pi^{2}\sqrt{2}m_{q}}\binom{C}{S} \Big\{\frac{1}{q^{2}}+\frac{p\cdot q}{q^{4}}-\frac{1}{3}\Big[\frac{p^{2}}{q^{4}}-\frac{4(p\cdot q)^{2}}{q^{6}}\Big]\Big\}+\frac{i\gamma_{5}f_{3\pi}}{4\pi^{2}\sqrt{2}} \binom{C}{S} \times \cr &&\Big\{p^{2}\log{(-q^{2})}+\frac{2(q\cdot p)^{2}}{q^{2}}-4p^{2}\log{(-q^{2})}-\frac{3p^{2}(q\cdot p)}{q^{2}}+\frac{2(q\cdot p)^{3}}{q^{4}}+\frac{4p^{2}(q\cdot p)}{q^{2}}\Big\}+ \sigma^{\alpha\beta}p_{\alpha}q_{\beta}\gamma_{5}\Big\{\frac{h_{q}}{12\sqrt{2}\pi^{2}m_{q}}\binom{C}{S}\times \cr &&\Big[-\frac{\log{(-q^{2})}}{4}+\frac{(p\cdot q)}{4q^{2}}-\frac{3}{10}\Big(\frac{p^{2}}{q^{2}}-\frac{2(p\cdot q)^{2}}{q^{4}}\Big)+\frac{1}{30}\Big(\frac{4(q\cdot p)^{3}}{q^{6}}-\frac{3p^{2}q\cdot p}{q^{4}}\Big)+ \frac{\langle g^{2}G^{2}\rangle}{288 q^{4}}\Big(1+\frac{2q\cdot p}{q^{2}}\Big)\Big]-\langle\overline{q}q\rangle\frac{2f_{q}}{3\sqrt{2}}\times \cr && \binom{C}{S}\Big[\frac{2}{q^{2}}+\frac{2 q\cdot p}{q^{4}}+\frac{10\delta^{2}}{9q^{4}}+\frac{20\delta^{2}p\cdot q}{9 q^{6}}\Big] + \frac{2f_{q}\delta^{2}\langle\overline{q}q\rangle}{9\sqrt{2}q^{4}}\binom{C}{S}\Big[1+\frac{2p\cdot q}{q^{2}}\Big] -\frac{\langle\overline{q}g\sigma\cdot Gq\rangle f_{q}}{4\sqrt{2}q^{4}}\binom{C}{S} \Big[1+\frac{2q\cdot p}{q^{2}}\Big]\Big\}.
\end{eqnarray}
\end{widetext}

\begin{widetext}
\par From Eqs. (\ref{eq7}) and (\ref{eq8}), we get
\begin{eqnarray}\label{eq15}
\frac{1}{\pi}\int dq_{0}^{\prime}e^{(-\frac{q_{0}^{\prime 2}}{M^{2}})}\hat{B}\text{Im}\Pi_{+}(q^{\prime},p) &=& \overline{u}(q)i\gamma_{5}u(q-p)\frac{\lambda^{2}}{E_{k}+\omega_{p}-M_{n}} \cr && \times  \Big\{e^{-\frac{M_{n}^{2}}{M^{2}}}g(M_{n},{\bf{p}}^{2})
-e^{-\frac{(E_{k}+\omega_{p})^{2}}{M^{2}}}g(E_{k}+\omega_{p},{\bf{p}}^{2})+\mathrm{cont.}\Big\}, \nonumber\\
\frac{1}{\pi}\int dq_{0}^{\prime} q_{0}^{\prime} e^{(-\frac{q_{0}^{\prime 2}}{M^{2}})}\hat{B}\text{Im}\Pi_{+}(q^{\prime},p) &= & \overline{u}(q)i\gamma_{5}u(q-p)\frac{\lambda^{2}}{E_{k}+\omega_{p}-M_{n}} \cr && \times \Big\{e^{-\frac{M_{n}^{2}}{M^{2}}}g(M_{n},{\bf{p}}^{2})M_{n}
-e^{-\frac{(E_{k}+\omega_{p})^{2}}{M^{2}}}(E_{k}+\omega_{p})g(E_{k}+\omega_{p},{\bf{p}}^{2})+\mathrm{cont.}\Big\},
\end{eqnarray}
where last terms stand for the continuum contributions.
Similarly, we can get Borel transform of OPE expression $\Pi(q,p)_{even}^{OPE}$ and $\Pi(q,p)_{odd}^{OPE}$. In the region of $M^{2}$ where sum rules work, we can equate the coefficients of $\overline{u}(q)i\gamma_{5}u(q-p)$ in the phenomenological expression Eq. (\ref{eq15}) to the coefficients of the same in OPE expressions and transfer the continuum contribution to the OPE side. The coefficient of $\overline{u}(q)i\gamma_{5}u(q-p)$ in the OPE expression for $\Pi_{+ odd}$ gives:
\begin{eqnarray}\label{eq16}
\hat{B}\left[\Pi^{OPE-cont.}_{+odd}\right] &=& M^{2}\Big\{\frac{f_{q}{\bf{p}}^{2}}{3\sqrt{2}\pi^{2}}\binom{C}{S}-\frac{a_{\eta^{(\prime)}}}{6\pi^{2}}\Big[3+\ln{4}-3\gamma+\ln {M^{2}}\Big]+\frac{h_{q}}{8\sqrt{2}\pi^{2}m_{q}}\binom{C}{S}\Big[\frac{5}{6}M_{n}-\frac{7}{6}E_{k}\Big]\Big\}E_{0}\Big(\frac{s_{\eta^{(\prime)}}}{M^{2}}\Big)  \cr && +\frac{h_{q}}{8\sqrt{2}\pi^{2}m_{q}} \binom{C}{S}\Big\{\frac{1}{3}(M_{n}-E_{k})^{2}\Big[2M_{n}+E_{k}\Big]+\frac{2}{5}E_{k}{\bf{p}}^{2}\Big\}-\frac{f_{3\pi}}{4\sqrt{2}\pi^{2}}\binom{C}{S}2(M_{n}-E_{k})^2
\Big\{2M_{n}-E_{k}\Big\} \cr &&
-\frac{4}{3\sqrt{2}}f_{q}\binom{C}{S}\langle\overline{q}q\rangle  \Big\{M_{n}+E_{k}\Big\}+\frac{f_{q}\delta^{2}}{6\sqrt{2}\pi^{2}}\binom{C}{S}\Big\{M_{n}-E_{k}\Big\}\frac{4}{3}E_{k} \cr && + \frac{1}{M^{2}}\Big\{\frac{1}{18\sqrt{2}}f_{q}\binom{C}{S}\langle\frac{\alpha_{s}}{\pi}G^{2}\rangle \Big[M_{n}-E_{k}\Big]\Big[M_{n}-5E_{k}\Big]+\frac{h_{q}\langle\overline{q}q\rangle}{9\sqrt{2}m_{q}}\binom{C}{S} \Big[M_{n}-E_{k}\Big]\Big[4E_{k}-2M_{n}\Big] \cr && - \frac{h_{q} \langle\frac{\alpha_{s}}{\pi}G^{2}\rangle}{48\times18\sqrt{2}m_{q}}\binom{C}{S}\Big[19M_{n}-17E_{k}\Big]+\frac{f_{q}}{\sqrt{2}}\binom{C}{S} \Big[M_{n}+E_{k}\Big]\Big[\frac{14}{27}\langle\overline{q}q\rangle\delta^{2}+\frac{1}{4}\langle\overline{q}g\sigma\cdot Gq\rangle\Big]\Big\},
\end{eqnarray}
\end{widetext}
where $E_{k}=\sqrt{M_{n}^{2}+{\bf{p}}^{2}}$, $\omega_{p}=\sqrt{m_{\eta^{(\prime)}}^{2}+{\bf{p}}^{2}}$. The continuum contribution has been parameterized in a standard way\cite{Ioffe:2010} through $E_{0}(x)$$=$$1-e^{-x}$. Unlike the case of $g_{\pi NN}$, here we have retained terms up to $O((M_{n}-E_{k})^{3})$.
 We have numerically checked that the short-distance expansion, as given in Eq.(14), works. We also use nucleon mass sum rule\cite{Ioffe:2010}:
\begin{eqnarray}\label{eq17}
M_{n}\lambda^{2}e^{-M_{n}^{2}/M^{2}}\cr &&=\frac{1}{4\pi^{2}}\{-M^{4}\langle\overline{q}q\rangle E_{1}(\frac{s_{0}}{M^{2}})+ \cr && \frac{\pi^{2}}{6}\langle\overline{q}q\rangle\langle\frac{\alpha_{s}}{\pi}G^{2}\rangle\}\equiv\hat{B}[\Pi^{m}].
\end{eqnarray}
At the physical point, where both the nucleon lines and the meson line are on mass-shell with ${\bf{p}}^{2}$=$ -m_{\eta^{(\prime)}}^{2}+\frac{m_{\eta^{(\prime)}}^{4}}{4 M_{n}^{2}}$, RHS of Eq. (\ref{eq7}), combined with Eq. (\ref{eq8}), takes $\frac{0}{0}$ form and hence can not be used to determine the coupling constants directly in this approach. However, we can determine $g_{\eta^{(\prime)} NN}$ at different kinematical points and extrapolate the results to the physical point.
Working with $\Pi_{+ odd}$, from the ratio of the two sum rules the coupling constant can be obtained as:
\begin{dmath}\label{eq18}
\footnotesize \frac{g(M_{n},{\bf{p}}^{2})}{M_{n}(E_{k}+\omega_{p}-M_{n})}=\frac{\hat{B}[\Pi_{+ odd}^{OPE-cont.}]}{\hat{B}[\Pi^{m}]} - \frac{M^{4}}{(E_{k}+\omega_{p})^{2}-M_{n}^{2}}\frac{d}{dM^{2}}(\frac{\hat{B}[\Pi_{+ odd}^{OPE-cont.}]}{\hat{B}[\Pi^{m}]}).
\end{dmath}
\begin{table*}
{\footnotesize \caption{Values of $g_{\eta NN}$ and $g_{\eta^{\prime} NN}$ obtained at different non-physical points and extrapolated at the physical point. Values of coupling constants due to gluonic operator $\alpha_{s}G\tilde{G}$, contribution without gluonic operator and s-quark contribution as it appears in $a_{\eta^{(\prime)}}$ are also shown.}\label{tab:coupling1}
\begin{ruledtabular}
\begin{tabular}{c c c c| c c c c}
\multicolumn{4}{c}{$g_{\eta NN}$ ($M^{2}$=$2 \text{GeV}^{2}$)}& \multicolumn{4}{c}{$g_{\eta^{\prime} NN}$  ($M^{2}$=$1.3 \text{GeV}^{2}$)}\\
\hline
${\bf{p}}^2$ & $-m_{\eta}^{2}/2$& $-\frac{3}{4}m_{\eta}^{2}$ & $-m_{\eta}^{2} + \frac{m_{\eta}^{4}}{4 M_{n}^{2}}$ & $-m_{\eta^{\prime}}^{2}/2$& $-2\frac{m_{\eta^{\prime}}^{2}}{3}$ &$-\frac{3}{4}m_{\eta^\prime}^{2}$ & $-m_{\eta^\prime}^{2} + \frac{m_{\eta^\prime}^{4}}{4 M_{n}^{2}}$\\
\hline
$g_{\text{total}}$			& 1.88  & 1.32 & 0.96   	 		& 3.07  & 1.69  &0.94 &1.03\\
$g_{\text{no-gluon}}$	    & 1.45  & 1.26  & 1.13        & 1.22  & 1.31  &1.44  & 1.42 \\
$g_{\text{gluon}}$		    & 0.43  & 0.06 & -0.17 		    & 1.85 &  0.38 &-0.50  & -0.39\\
$g_{\text{s}}$				& 0.82 & 0.12 & -0.34 			& 1.11 &     0.23 & -0.3 & -0.24\\
\end{tabular}
\end{ruledtabular}}
\end{table*}
Following values of parameters have been used for estimation of $g_{\eta^{(\prime)} NN}$ (all quantities are in  GeV unit) \cite{Kim:2000,Kondo:2003,Singh:2011,Ioffe:2010,Singh:2015}:
$\langle\overline{q}q\rangle$=$-(1.65\pm0.15)\times 10^{-2}$, $\langle \frac{\alpha_{s}}{\pi} G^{2}\rangle $=$0.005\pm0.004$, $\delta^{2}$=$0.2\pm0.04$, $f_{q}$=$(1.07\pm0.02)f_{\pi}$, $f_{s}$=$(1.34\pm0.06)f_{\pi}$, $s_{\eta^{(\prime)}}$=$2.57\pm0.03$, $s_{0}$=2.5, $f_{3\pi}$=0.0045, $\phi$=$40\degree\pm1\degree$, $\frac{h_{q}}{m_{q}}=-4\frac{f_{q}}{f_{\pi}^{2}}\langle\overline{q}q\rangle$ \cite{Ball:2007}, $\langle\overline{q}g\sigma\cdot Gq\rangle$=$m_{0}^{2}\langle\overline{q}q\rangle$, $m_{0}^{2}$=$0.8\pm0.1$.
\begin{table*}
{\footnotesize \caption{Comparison of our results on $g_{\eta NN}$ and $g_{\eta^{\prime} NN}$ with results for the same from recent literature.}\label{tab:coupling2}
\resizebox{\textwidth}{!}{
\begin{ruledtabular}
\begin{tabular}{c c c c c c c c}
Ref. & $g_{\eta NN}$& $g_{\eta^{\prime} NN}$ & Comment\\
\hline
Present work & $(0.64-1.26)$ & $(0.44-1.27)$ & QCD sum rule\\
\cite{Nasrallah:2007}  & $(4.95-5.45)$ & $(5.6-10.9)$ & GT relation+ Dispersion relation \\
\cite{Feldmann:2000}   &  $(3.4\pm0.5)$ & $(1.4\pm1.1)$ & Theory (GT relation) \\
\cite{Nys:2017}		  & $2.241$      &$-$  & Photoproduction \\
\cite{Shore:2006}     & $(3.78\pm0.34)$ & $(1-2)$ & Theory\\
\cite{Aznauryan:2003} & $(0.39,0.92)$ &$-$ & Photoproduction+(Isobar model, Dispersion relation)\\
\cite{Tiator:2018}    & $0.89$  & $0.87$   & Fitting photoproduction data\\
\cite{Singh:2011}     & $4.2\pm1.05$ &$-$ & QCDSR at unphysical point\\
\cite{Yang:2018}      & $4.399\pm0.365$ & $2.166\pm0.312$ & Chiral quark-soliton model\\
\cite{Rijkin:2010}	  & $6.852$ & $8.66$  & Potential model\\
\end{tabular}
\end{ruledtabular}}}
\end{table*}
\section{Analysis of Sum Rules and Results}
We have plotted the coupling constant g obtained from Eq. (\ref{eq18}) in Figs. (2) and (3) as a function of $M^{2}$ for different values of ${\bf{p}}^{2}$. Our chosen range of the Borel mass is $0.8$ \text{GeV}$^{2}<M^{2}<1.8$ \text{GeV}$^{2}$ for $\eta^{\prime}$ and $1.5$ \text{GeV}$^{2}<M^{2}<2.5$ \text{GeV}$^{2}$ for $\eta$. This is done with an eye to keep continuum contribution and contribution of $1/M^{2}$ terms less than 30 \% and to keep $g_{\eta^{\prime} NN}$ not too low. A similar analysis with the even correlation function $\Pi^{E}_{+}$ gives large continuum contribution and the large contribution of $1/M^{2}$ terms, hence the result is not reliable.  In Table-1, we have displayed our results for coupling constants, coupling constants without gluonic contribution, the gluonic contribution to the coupling constants and the contribution coming from the OZI-rule violating s-quark ($f_{s}$ term in $a_{\eta^{(\prime)}}$, see Eq. (\ref{eq13})). It is clear from the Table that though the gluonic contribution and the OZI rule violating contribution from s-quark are small at the physical point, they become significantly large off the physical point and eventually become dominant at far off physical point for $\eta^{\prime}$. These coupling constants off the physical point will be important for some processes such as the photo-production of mesons off a nucleon target. We have made error estimates of our results as follows: Errors due to different phenomenological parameters, as given above, and error due to finite slope and finite range of the Borel mass have been shown separately. Error due to deviation from linear extrapolation is small and is neglected.
\begin{eqnarray}\label{eq19}
g_{\eta NN} &=& 0.96^{+ 0.16}_{-0.17} (\text{M}^{2})^{+0.14}_{-0.15}(\text{rest}), \nonumber \\
g_{\eta^{\prime} NN} &=& 0.76^{+ 0.27}_{-0.08} (\text{M}^{2})^{+0.24}_{-0.24}(\text{rest}).
\end{eqnarray}
The flavor-singlet Goldberger-Treimann relation for QCD was derived by Shore and Veneziano\cite{Shore:1992},
\begin{dmath}\label{eq20}
M_{n}g_{A}^{(0)}=\sqrt{\frac{3}{2}}[f^{0\eta^{\prime}}(g_{\eta^{\prime}NN}+g_{gluon}^{(\eta^{\prime})})+f^{0\eta}(g_{\eta NN}+g_{gluon}^{(\eta)})]
\end{dmath}
Using our results for $g_{\eta^{(\prime)}NN}$ and $g_{gluon}^{(\eta^{(\prime)})}$, and $f^{0\eta^{(\prime)}}$ from Ref.\cite{Shore:2006}, we get $g_{A}^{(0)}=(0.23-0.28)$ in the range of medium to maximum values of the parameters. This may be compared with the results obtained by COMPASS($Q^{2}=3\text{GeV}^{2}$)\cite{Adolph:2016}:$g_{A}^{(0)}$=$[0.26-0.36]$ and NNPDFpoll.1($Q^{2}=10 \text{GeV}^{2}$)\cite{Nocera:2014}: $g_{A}^{(0)}=0.25\pm0.10$ and a theoretical estimate \cite{Singh:2015}:$g_{A}^{(0)}$=$0.39\pm0.05\pm0.04$, though it is smaller than the lattice results\cite{Liang}:$g_{A}^{(0)}$=0.405.
\begin{figure}[H]
\includegraphics[width=0.3\textwidth]{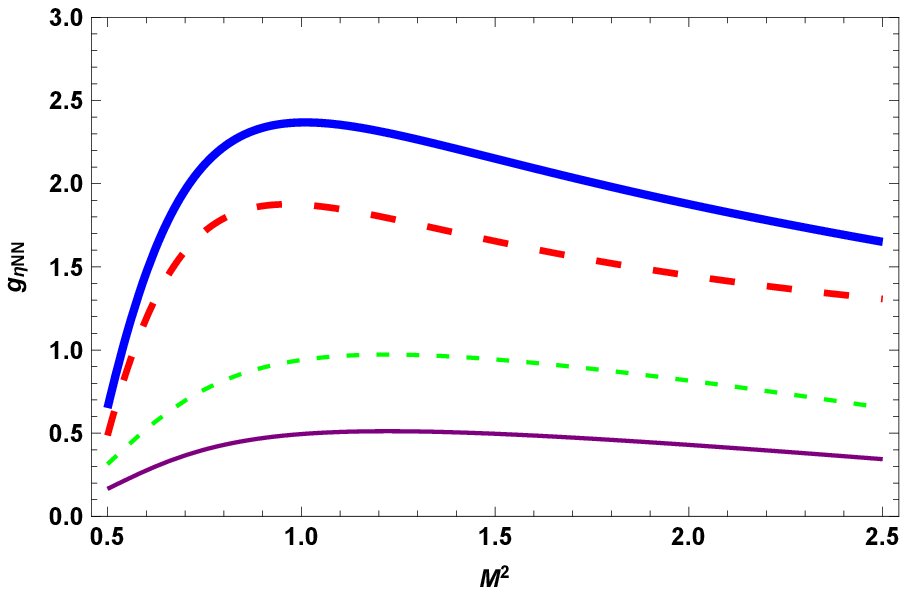}{(a)}\\
\hfill\hfill\includegraphics[width=0.3\textwidth]{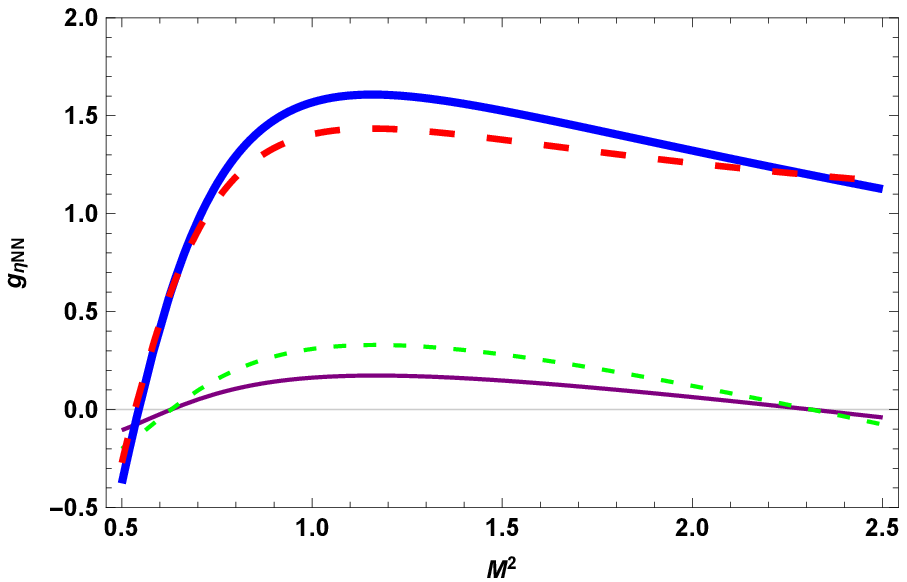}{(b)}
\caption{\footnotesize Plots of our results for $g_{\eta NN}$  as a function of $M^{2}$(thick solid line). Also plotted are our results on $g_{\eta NN}$ without contribution from gluonic operator $\alpha_{s}G\tilde{G}$ (long-dashed line),only due to $\alpha_{s}G\tilde{G}$(thin solid line) and due to the part with $f_{s}$ in $a_{\eta}$(short-dashed line). Figs.2(a) and 2(b) are for $\eta$ when ${\bf{p}}^{2}$  is -$\frac{ m_{\eta}^{2}}{2}$ and -$\frac{3 m_{\eta}^{2}}{4}$ respectively.}
\end{figure}
\begin{figure}[H]
\includegraphics[width=0.3\textwidth]{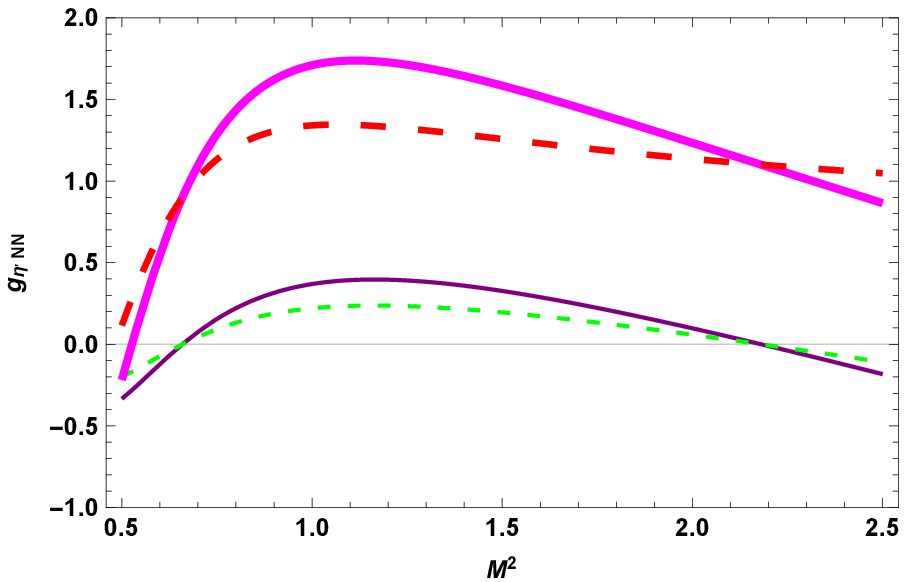}{(a)}\\
\hfill\hfill\includegraphics[width=0.3\textwidth]{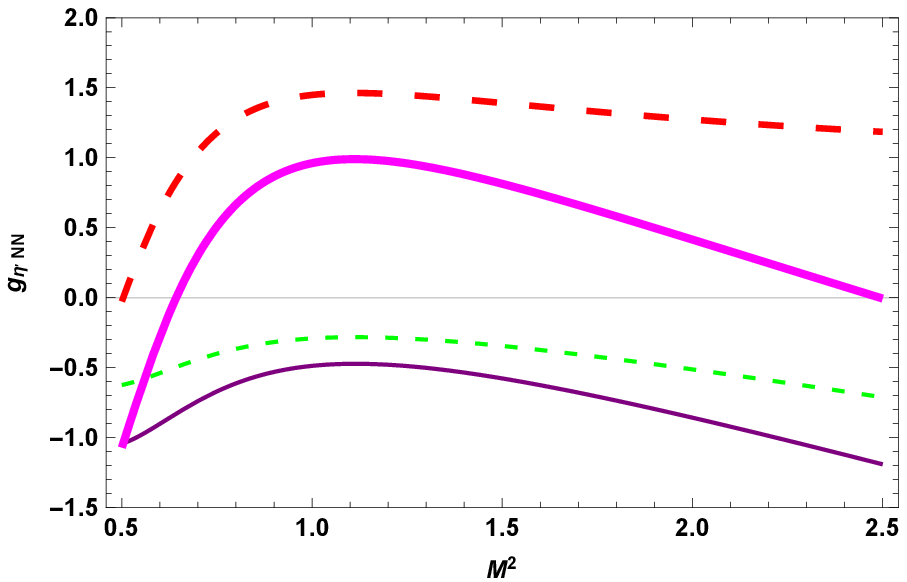}{(b)}
\caption{\footnotesize Plots of our results for  $g_{\eta^{\prime} NN}$ as a function of $M^{2}$(thick solid line). Also plotted are our results on  $g_{\eta^{\prime} NN}$ without contribution from gluonic operator $\alpha_{s}G\tilde{G}$ (long-dashed line),only due to $\alpha_{s}G\tilde{G}$(thin solid line) and due to the part with $f_{s}$ in $a_{\eta^{\prime}}$(short-dashed line).  Figs.3(a) and 3(b) are for $\eta^{\prime}$ when ${\bf{p}}^{2}$  is -$\frac{2 m_{\eta^{\prime}}^{2}}{3}$ and -$\frac{3 m_{\eta^{\prime}}^{2}}{4}$ respectively.}
\end{figure}
\section{Summary and Conclusion}
We have used QCD sum rules, a well-tested approach in hadron physics, to calculate $g_{\eta NN}$ and $g_{\eta^{\prime} NN}$.The flavor-singlet contribution has been facilitated by use of quark-flavor basis. A characteristic contribution from radiatively generated gluonic operator $\alpha_{s}G\tilde{G}$ was explicitly included. Finally, the values of the coupling constants at the physical point were estimated from linear extrapolation of results obtained at two other points. Anomalous glue, which gives excess mass to the would-be Goldstone bosons $\eta$ and $\eta^{\prime}$ on the one end and a fraction of spin to the nucleon on the other end, also gives substantial contribution to the coupling constants of these mesons to the nucleon. Radiatively generated gluons, which eventually go to the mesons, are attached to the valance d-quark for the proton and to the valance u-quark for the neutron. Though interpolating fields of nucleons consist of u- and d-quark fields only, the coupling constants of the nucleon with $\eta$ and $\eta^{\prime}$ contain a term proportional to $f_{s}$. This can be considered as OZI-violating contribution to the meson-nucleon coupling constant and is substantial for both $\eta$ and $\eta^{\prime}$ mesons. Though $\eta$ is largely octet, the gluon contribution to the $g_{\eta NN}$ turns out to be significant. The branching ratios for N*(1535) to decay to $\eta N$ and $\pi N$ final states are approximately equal, about 45\%, even though the latter channel has larger phase space. The result is interpreted as evidence for a possible gluon anomaly contribution to the decay by Olbrich et al \cite{Olbrich}. Understanding of non-perturbative gluon dynamics and axial $U(1)$ anomaly has a vital role in future development of hadron physics and nuclear physics and our results on  $g_{\eta NN}$ and $g_{\eta^{\prime} NN}$ will be useful for this.
\begin{acknowledgments}
SDP acknowledges financial support from UGC-BSR.
\end{acknowledgments}

\end{document}